\documentclass[10pt, nobalancelastpage, twocolumn, floatfix, superscriptaddress, aps, prl]{revtex4-2}
\usepackage{amsmath, amssymb, graphicx, hyperref, xcolor, microtype}
\usepackage[export]{adjustbox}
\usepackage{braket}
\usepackage{enumitem}

\newcommand{\CNOT}{\mathit{CNOT}}

\begin{document}

\title{Automated quantum circuit optimization with randomized replacements}

\author{Marcin Szyniszewski}
\affiliation{Department of Computer Science, University of Oxford, Parks Road, Oxford OX1 3QD, United Kingdom}

\author{Aleks Kissinger}
\affiliation{Department of Computer Science, University of Oxford, Parks Road, Oxford OX1 3QD, United Kingdom}

\author{Noah Linden}
\affiliation{School of Mathematics, University of Bristol, Woodland Road, Bristol BS8 1UG, United Kingdom}

\author{Paul Skrzypczyk}
\affiliation{H. H. Wills Physics Laboratory, University of Bristol, Tyndall Avenue, Bristol BS8 1TL, United Kingdom}

\begin{abstract}
    Quantum circuit optimization -- the process of transforming a quantum circuit into an equivalent one with reduced time and space requirements -- is crucial for maximizing the utility of current and near-future quantum devices. While most automated optimization techniques focus on transforming circuits into equivalent ones that implement the same unitary, we show that substantial new opportunities for resource reduction can be achieved by (1)~allowing approximate local transformations and (2)~employing mixed quantum channels to approximate pure circuits. Our novel automated protocol for approximate circuit rewriting is a refined evolution of automated optimization techniques based on the ZX-calculus, where we add a greedy strategy that selectively replaces ZX-diagrams with small phase angles with stochastic mixtures of the identity and carefully chosen over-rotations, which are designed to reduce the overall gate count in expectation while staying within a strict error budget. This approach yields modest two-qubit gate count reduction in random quantum circuits, and achieves a substantial reduction in structured circuits such as the quantum Fourier transform. Fundamentally, our protocol converts experimental noise due to gate applications into deliberately engineered random noise, outperforming many other approximation methods on average. These results highlight the potential of mixed-channel approximations to enhance future quantum circuit performance, suggesting new directions for resource-aware automated quantum compilation beyond pure unitary channels.
\end{abstract}

\maketitle

\textbf{\textit{Introduction.}}---%
The increasing complexity of quantum algorithms, alongside persistent constraints in current quantum hardware, has elevated circuit optimization from a technical convenience to a practical necessity. While current and near-future quantum computers are capable of executing non-trivial computations, they are severely hampered by short decoherence times and gate noise, which puts major limitations on circuit depth and gate count~\cite{Preskill2018, Wilson2021}. Quantum error correction and full fault-tolerant computation offer a long-term remedy for physical gate noise, but at the cost of high time and space overheads to store and perform basic operations on encoded qubits~\cite{Fowler2012, Litinski2019}. Consequently, substantial effort has been directed toward automated circuit optimization techniques that minimize circuit costs.

Historically, these optimization methods have often prioritized exactness, transforming a quantum circuit into one that implements the same unitary through local gate identities~\cite{Nam2018}, algebraic transformations such as phase polynomial reduction~\cite{Nam2018, Amy2013}, or graph-theoretic techniques based on the ZX-calculus~\cite{Duncan2020, Staudacher2023}. However, for many quantum applications, such as any algorithm that produces the correct answer with success bounded above 1/2,
exact reproduction of the target unitary is not strictly required. Approximations within known error bounds are often acceptable, and in practice, quantum circuits are frequently approximate to begin with, for example via Trotterisation~\cite{Campbell2019, Childs2021, Morales2025} or compilation to a finite gate set like Clifford+T~\cite{Kitaev1997, Dawson2006, Kliuchnikov2013, Kliuchnikov2023, Selinger2015}. It is therefore natural to consider approximate circuit optimization methods, where small fidelity loss can yield substantial savings in two-qubit gate count, circuit depth, or overall execution time.

\begin{figure}[b]
  \centering
  \includegraphics[width=\columnwidth]{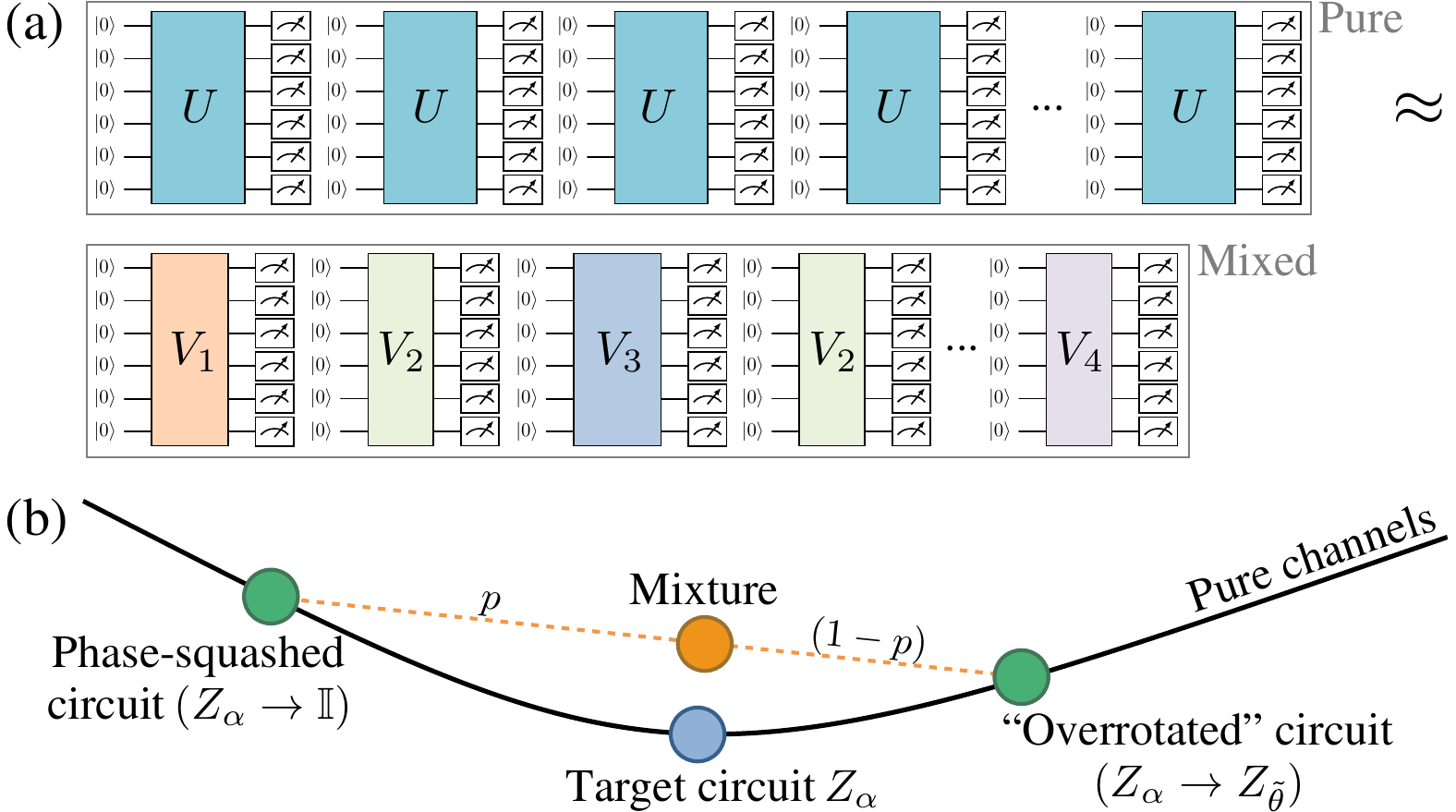}
  \caption{(a)~Circuit $U$ run repeatedly on a quantum computer (top) can be approximated using a mixture of slightly different circuits $V_n$ (bottom), which reduces average gate count. (b)~Intuitive picture showing how a circuit containing $Z$-phase gate $Z_\alpha$ can be approximated by a mixture of phase-squashed circuit ($Z_\alpha$ replaced by identity) with probability $p$ and an ``overrotated'' circuit with $Z_{\tilde{\theta}}$ ($\tilde{\theta} > \alpha$), with probability $(1-p)$. \label{fig:summary}}
\end{figure}

In this landscape, the ZX-calculus offers some advantages. Its graph-theoretic representation absorbs many common circuit identities, such as obvious gate commutations, and as was recently shown by one of the authors~\cite{Kelly2025}, the types of ZX-diagrams most commonly used in quantum circuit optimization are robust, in the sense that small local changes of parameters only cause small global errors in the diamond distance.
In Ref.~\cite{Kelly2025}, the authors used local transformations that round small-angle phases in ZX-diagrams to the nearest \textit{Clifford angle} (i.e., the nearest integer multiple of $\pi/2$). This, in turn, enables more optimizations using the ZX-calculus, where Clifford angles play a special role. This not only provides precise control over the induced error but also demonstrates that the approximation step is amenable to full automation, thereby removing the need for user oversight.

In this work, we introduce a new automated approximation protocol via stochastic mixing that offers a more resource-efficient alternative to algorithms restricted to pure circuits. In general, allowing for mixed channels can yield superior approximations~\cite{Campbell2017, Hastings2017} with reduced gate counts, under a fixed error budget. In realistic quantum computing setups, measurement data is obtained by averaging over many repeated executions (shots) of the same circuit. However, due to inherent experimental noise, the circuit does not implement an identical unitary each time, but rather behaves as a mixed channel. Furthermore, since stochastic randomization can also lower the resource requirements of a circuit~\cite{Campbell2019, Hashim2021, Wan2022, Kliuchnikov2023, Koczor2024, Akibue2024}, and hence physical noise, one can see quantum circuit approximation as trading one kind of noise for another. 
Motivated by these observations, we propose exploiting the necessity of multiple shots to approximate a pure unitary circuit by a mixed channel, where each shot in principle corresponds to a slightly different (and often shorter) circuit [see Fig.~\ref{fig:summary}(a)]. Our protocol thus fits naturally into the current ``multi-shot'' paradigm used by today's quantum devices while enabling substantial savings in the expected circuit size, averaged across all of the shots.

\textbf{\textit{Methodology.}}---%
We begin by introducing our core approximation strategy: the local replacement of individual unitary gates with mixed-channel alternatives. We rigorously assess the fidelity of such replacements by employing the diamond distance as our primary error metric (albeit other metrics often give consistent results), leveraging its compositional properties to bound cumulative errors across the circuit. Finally, we detail our automated approximation protocol, which involves iteratively applying local replacements and optimizing using ZX-calculus.

Any quantum circuit can be represented using a universal set of gates, consisting of $\CNOT$, Hadamard $H$, phase gates $S$, and $Z$-phase gates $Z_{\alpha}$, where
$Z_{\alpha} = \left(\begin{smallmatrix} 
    1 & 0\\
    0 & e^{i \alpha}
  \end{smallmatrix}\right)$.
The set $\{\CNOT, H, S\}$ generates all Clifford gates. We can restrict $\alpha$ to the range $(- \pi / 4, \pi / 4]$ without loss of generality since any $Z_\beta$ ($\beta$ arbitrary) equals a sequence of $S$ gates followed by $Z_\alpha$. This definition also ensures that the nearest Clifford gate to $Z_\alpha$ is always a $2 \times 2$ identity matrix $\mathbb{I}$. In fact, when $\alpha$ is small, $Z_\alpha$ is well approximated by $\mathbb{I}$, motivating an approximation technique that replaces $Z_\alpha$ by $\mathbb{I}$. Such a protocol (introduced in Ref.~\cite{Kelly2025} and dubbed ``phase squashing'') often allows for further circuit optimization, reducing the number of two-qubit gates -- for example, if $Z_\alpha$ appeared between $U$ and $U^{-1}$, after squashing, $UU^{-1}$ cancels completely. Note that a similar idea is used in approximate quantum Fourier transform~\cite{Cleve2000, Ahokas2004} (aQFT), reducing the gate count by a factor of $\log L/L$, where $L$ is the number of qubits.

Now, we inject mixing into this framework. An intuitive illustration of the potential benefits of mixing is provided in Fig.~\ref{fig:summary}(b): the mixture of two pure channels can induce a smaller approximation error (represented as Euclidean distance in the figure) than phase-squashing alone. Specifically, the ``underrotated'' circuit is chosen optimally to be the phase-squashed circuit ($Z_\alpha \to \mathbb{I}$), which reduces the number of two-qubit gates, while the ``overrotated'' circuit, $Z_\alpha \to Z_\theta$ ($\theta>\alpha$), retains the original gate count. Formally, we define two quantum channels: a pure channel $\mathcal{Z}_{\alpha}(\rho)$ implementing the target $Z$-phase gate $Z_{\alpha}$,
\begin{align}
  \mathcal{Z}_{\alpha} (\rho) &= Z_{\alpha} \rho Z_{\alpha}^{\dagger} \label{eq:channel_original} \\
  &= \cos^2 (\tfrac{\alpha}{2}) \rho + \sin^2 (\tfrac{\alpha}{2}) Z \rho Z 
  - i \sin (\tfrac{\alpha}{2}) \cos (\tfrac{\alpha}{2}) [Z,
  \rho],\nonumber
\end{align}
and a mixed channel $\mathcal{E}_{\theta, p}(\rho)$, containing a mixture of $\mathbb{I}$ with probability $p$ and $Z_{\theta}$ with probability $(1 - p)$,
\begin{align}
    \mathcal{E}_{\theta, p} (\rho) &= p \rho + (1 - p) Z_{\theta} \rho Z_{\theta}^{\dagger}\label{eq:channel_approximate}\\
    & = p \rho + (1 - p) \cos^2 (\theta / 2) \rho + (1 - p) \sin^2 (\theta / 2) Z \rho Z \nonumber \\
    & \quad  - (1 - p) i \sin (\theta / 2) \cos (\theta / 2) [Z, \rho] . \nonumber
\end{align}
When $\theta \approx \alpha/(1-p)$, the channels become approximately equivalent to leading order in $\alpha$, but the mixture reduces the average gate count.

We quantify the difference between two channels using the diamond distance, which is endowed with certain useful properties, such as the triangle inequality, and subadditivity w.r.t.\@ tensor products. Hence, the diamond distance of many replacements (on potentially different qubits) is bounded from above by the sum of the diamond distances of each single replacement. Diamond distance between two channels $\mathcal{E}_1$ and $\mathcal{E}_2$, which operate on Hilbert space $\mathcal{X}$, is defined as
\begin{equation}
  d_{\diamond} (\mathcal{E}_1, \mathcal{E}_2) = \max_{\substack{M \in L (\mathcal{X} \otimes
  \mathcal{X}),\\ \| M \|_1 < 1}} \| (\mathcal{E}_1 \otimes \mathbb{I}_{\mathcal{X}})
  (M) - (\mathcal{E}_2 \otimes \mathbb{I}_{\mathcal{X}}) (M) \|_1,
\end{equation}
where $M$ is a linear map operating on $\mathcal{X} \otimes \mathcal{X}$, $\mathbb{I}_{\mathcal{X}}$ is the identity on $\mathcal{X}$, and $\| \cdot \|_1$ is the trace norm. The diamond distance measures the maximum possible operational difference between two channels given the same initial state, i.e., distinguishability in the worst case.

We derive the formula for the diamond distance of one $Z$-phase gate replacement (see End Matter), given by
\begin{align}
  d_{\diamond} (\mathcal{Z}_{\alpha}, \mathcal{E}_{\theta, p}) 
  &= \sqrt{2}  \Big((p - 1) \sin \alpha \sin
  \theta  +
  p^2 - p + 1 \label{eq:dist} \\
  &\quad + (p - 1) (\cos \alpha - p) \cos \theta - p \cos \alpha \Big)^{1/2} .
  \nonumber
\end{align}
Then, the optimal value of $\theta$, denoted $\tilde{\theta}$, is given by
\begin{equation}
    \tilde{\theta} = 2 \tan^{- 1} \!\left(\!\frac{p - \cos (\alpha) + \sqrt{1 + p^2 - 2 p
  \cos (\alpha)}}{\sin (\alpha)} \right)\!,
  \label{eq:thetamin}
\end{equation}
for which the diamond distance assumes the lowest value,
\begin{align}
    d_\diamond (\mathcal{Z}_{\alpha}, \mathcal{E}_{\tilde{\theta}, p}) &= \sqrt{2} \Big( 1-p+p^2-p \cos \alpha \label{eq:diamondmin}\\
    &\quad-(1-p) \sqrt{1+p^2-2p \cos \alpha}\Big)^{1/2}.\nonumber
\end{align}
For small $\alpha$, and $p$ bounded away from 1, these can be Taylor-expanded as
\begin{equation}
    \tilde{\theta} = \frac{\alpha}{1 {-} p} +O(\alpha^3), \quad
    d_\diamond (\mathcal{Z}_{\alpha}, \mathcal{E}_{\tilde{\theta}, p}) = \frac{p
  \alpha^2}{2 (1 {-} p)}+O(\alpha^4). \label{eq:approxthetamin}
\end{equation}

The diamond distance, while a widely adopted metric in quantum information, tends to overestimate errors in practical settings (e.g., when evaluating experimental architectures), as it captures the worst-case error between channels -- often corresponding to outlier circuits that are rarely encountered in practice. We have therefore also considered typical/average distance measures~\cite{Maciejewski2023ieee, Maciejewski2023quantum} (see End Matter). Interestingly, for a single $Z_\alpha$ gate replacement, several commonly used typical distances scale proportionally to the diamond distance Eq.~\eqref{eq:dist}, and hence, computing the diamond distance not only provides a rigorous upper bound on the worst-case error, but also yields meaningful estimates of typical errors, making it (in our case) relevant in experimental settings.

To simplify the circuit after a replacement, we use the ZX-calculus, a graphical language for reasoning about quantum circuits (and in general, linear maps). Any circuit can be written as a ZX-diagram -- a graph representing a tensor network. 
The ZX-diagrams can be deformed using a set of graphical rewrite rules~\cite{Vilmart2019}, 
which are the main tool for simplification purposes. After simplification, the circuit extraction from the ZX-diagram can be done efficiently in polynomial time due to the preservation of a property called generalized flow~\cite{Kelly2025}.
For numerical tests, we use PyZX~\cite{Kissinger2020PyZX}, a Python package that implements simplification rules of the ZX-calculus. PyZX is generally optimized to minimize the T-gate count~\cite{Kissinger2019}, while the two-qubit gate count is of secondary importance. Hence, we specifically try two simplification algorithms (\texttt{simplify.basic\_simp} and \texttt{full\_reduce}), and choose the result with the lowest two-qubit gate count.

We now detail the approximation protocol, where a unitary circuit is approximated by a quantum channel.
\setlist{nolistsep}
\begin{enumerate}[label={\textbf{(\arabic*)}}, leftmargin=*, align=left]
    \item \textbf{Initialization:} Choose an error budget $\varepsilon$ and $p$. \label{step_setup}
    
    \item \textbf{Preprocessing:} \label{step_preprocessing}

    \begin{enumerate}[leftmargin=*, align=left]
        \item Sort all $Z_\alpha$ gates by the diamond distances from Eq.~\eqref{eq:diamondmin} for their replacements by mixtures. \label{step_calc_dist}
        
        \item Starting from $Z_\alpha$ with the lowest diamond distance, test whether replacing $Z_\alpha \to \mathbb{I}$ reduces the two-qubit gate count using ZX-calculus simplification. If yes, the replacement is added to the list of accepted replacements. Repeat until the accumulated diamond distance of all accepted replacements reaches $\varepsilon$.~\footnote{Note that Step~\ref{step_accept_repl} mimics the ``greedy algorithm'' from Ref.~\cite{Kelly2025}.} \label{step_accept_repl}
    \end{enumerate}

    \item \textbf{Implementation:} For each run of the channel, stochastically replace each $Z_\alpha$ from the list of accepted replacements [$Z_\alpha \to \mathbb{I}$ with probability $p$ and $Z_\alpha \to Z_{\tilde{\theta}}$ with probability $(1-p)$], and simplify using ZX-calculus. \label{step_implementation}
\end{enumerate}
One may regard this protocol as introducing engineered noise: by reducing the two-qubit gate count, we effectively trade experimental noise (i.e.\@ that arises when performing a gate) for this artificial counterpart. Consequently, in a practical setting, this engineered noise should be treated in the same way -- by performing a sufficient number of shots to account for the statistical uncertainty.
Moreover, step~\ref{step_implementation} admits a statistical interpretation: each accepted replacement corresponds to a Bernoulli random variable, and each circuit instance is drawn by sampling these variables, resulting in a random unitary circuit where $Z_\alpha$ gate application is conditioned on the sampled outcomes. Also note that the number of accepted replacements is determined by the distribution of phase angles and $\varepsilon$, rather than by the system size or circuit depth.

\begin{figure*}
    \centering
    \includegraphics[width=\linewidth]{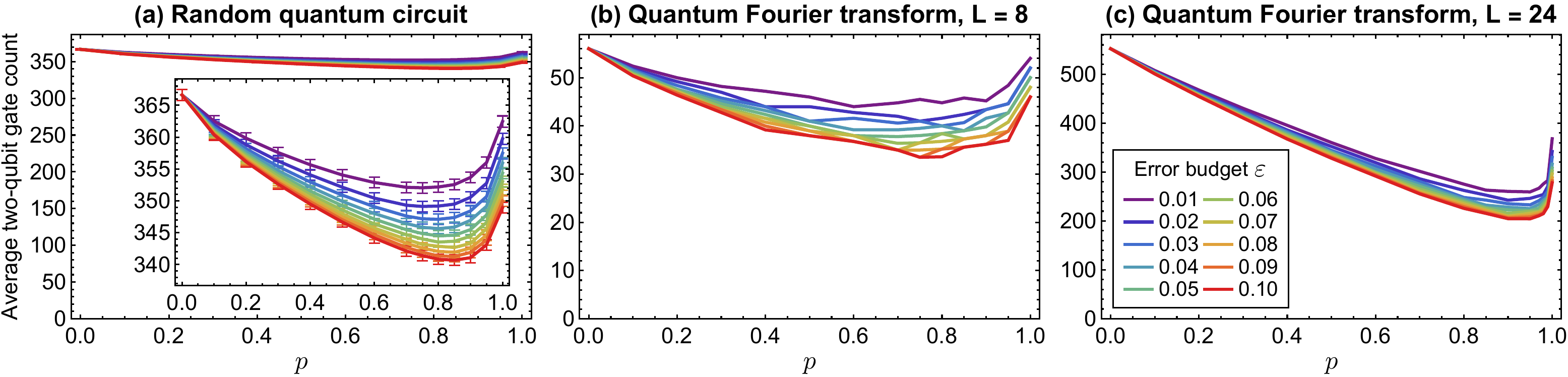}
    \caption{Average two-qubit gate count results of the approximation protocol for (a)~random quantum circuit for $L = 8$ qubits, (b)~quantum Fourier transform for $L=8$ qubits, and (c)~quantum Fourier transform for $L=24$ qubits, all as functions of identity replacement probability $p$, and under error budget $\varepsilon$. The inset in (a) shows a zoomed-in section of the plot. The legend in (c) applies in (a) and (b). \label{fig:results_gate_count}}
\end{figure*}

\textbf{\textit{Numerical results.}}---%
We perform the approximation protocol for a range of example circuits often found in the literature. We have chosen two specific scenarios representing the worst-case and best-case circuits for our approximation purposes.

First, we use a random quantum circuit (RQC), where optimization is expected to be challenging. Intuitively, this stems from the fact that in such circuits, replacing a $Z_\alpha$ gate with the identity rarely results in simplification -- the surrounding (randomly placed) gates are unlikely to be in a configuration that admits further reduction.
In detail, we use a random quantum circuit of depth $T$, where in each time step one randomly chooses to apply either $\CNOT$, $H$, $S$, or $Z_{\alpha}$, with a phase chosen uniformly on the interval $\alpha \in (- \pi / 4, \pi / 4]$. We fix the probabilities to be $p_{\CNOT} = 0.5, p_H = 0.3, p_S = 0.1, p_{Z_\alpha} = 0.1$, which is a generic choice that results in a relatively large number of $\CNOT$ gates, and thus large connectivity in the circuit. We will use $L = 8$ qubits, and the number of shots of $N_{\text{shots}} = 8192$, which is a typical shot count for an IBM quantum processor~\cite{IBMdocs}. Our results are averaged over $N_{\text{real}} = 1000$ random circuit realizations.

\begin{figure}[t]
  \centering
  \includegraphics[width=\columnwidth]{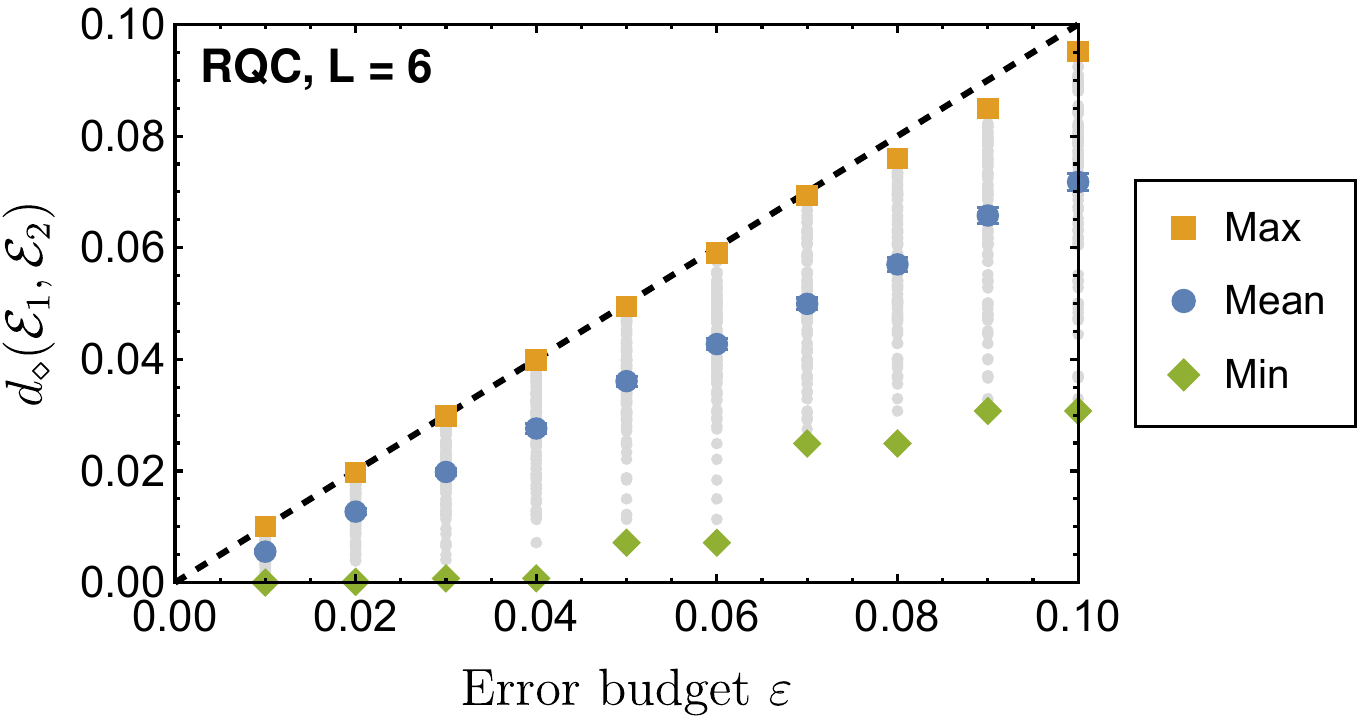}
  \caption{Diamond distance for the random quantum circuit ($L = 6$, $N_\text{real} = 100$, $p = 0.8$) as a function of error budget $\varepsilon$. All realizations are shown in light gray, their mean is in blue, the maximum in orange, and the minimum in green. The dashed line shows the error budget.\label{fig:diamond_distance}}
\end{figure}

The gate count results are presented in Fig.~\ref{fig:results_gate_count}, panel (a) showing those obtained for the random quantum circuit. For a fixed $\varepsilon$, we observe that the original gate count ($p = 0$) decreases when $p > 0$, reaching a minimum at $p \approx 0.8$. Our protocol also improves upon the phase squashing~\cite{Kelly2025}, which corresponds to $p = 1$. Nevertheless, the relative improvement in the gate count remains modest (below 10\%), as expected for random circuits: due to their structure, the error budget is relatively quickly saturated, and the number of replacements that could simplify the circuit is low. 
Numerical checks further confirm (see Fig.~\ref{fig:diamond_distance}) that the diamond distance $d_\diamond$ of the entire circuit remains within the error budget $\varepsilon$, while an example of a typical distance measure (Frobenius distance) remains approximately proportional to $d_\diamond$ (see End Matter).

Our second example is that of the quantum Fourier transform (QFT). Here, the improvements in the approximation protocols are expected to be much higher, as QFT includes a large number of $Z_{\alpha}$ gates with exponentially small phases, arranged in a structured way, a feature that is crucial in many aQFT protocols~\cite{Cleve2000, Ahokas2004}.
In detail, QFT can be written as the following unitary circuit,
\begin{equation}
    U_\text{QFT} = \prod_{i=0}^{L-1} \Big(\prod_{j=i+1}^{L-1} \mathit{CP}^{i,j}_{\pi/2^{j-i}}\Big) H^i,
\end{equation}
where top indices on gate names designate qubits, and $\mathit{CP}^{i,j}_\alpha$ is a controlled phase gate with target $i$, control $j$, and phase $\alpha$, and can be rewritten as $\mathit{CP}^{i,j}_{\alpha} = Z^j_{\alpha/2} \, \CNOT^{j,i} \, Z^i_{-\alpha/2} \, \CNOT^{j,i} \, Z^i_{\alpha/2}$.

We present results for QFT with $L = 8$ qubits in Fig.~\ref{fig:results_gate_count}(b) and $L = 24$ qubits in Fig.~\ref{fig:results_gate_count}(c). When $p$ is chosen optimally, the average two-qubit gate count is lowered significantly: for an error budget of $\varepsilon=0.1$, the gate count drops from 56 to $\approx 33$ for $L=8$, and from $552$ to $\approx 204$ for $L=24$. This represents a 41\% and 63\% relative drop. The optimal $p$, for which the gate count is minimal, varies from $p\approx0.75$ for $L=8$ to $p\approx0.93$ for $L=24$, suggesting that for larger system sizes, the optimal $p$ is just below $1$.

Our results should be compared to a well-established aQFT protocol~\cite{Cleve2000, Ahokas2004}, in which gates with angles below a certain threshold are dropped (equivalent to the $p = 1$ phase squashing). In fact, our protocol may be viewed as a mixed-channel generalization of this aQFT protocol. Both should exhibit the same asymptotic scaling in the gate count with the system size $(\sim L \log L)$; however, our protocol yields improvements in raw numbers. In practice, since the optimal $p$ is just below $1$, this means that instead of using the usual aQFT protocol, the effective strategy uses a mixture of mostly the original aQFT with even more dropped gates, plus a small number of circuits where some of the dropped gates are restored, but overrotated. Hence, our algorithm may occasionally produce longer circuits than aQFT; however, on average, we observe a significant reduction in gate counts (e.g., for $\varepsilon = 0.01$, a mixture allows us to drop 10 additional gates on average for $L=8$ QFT and 109 additional gates for $L=24$ QFT, compared to aQFT).

\textbf{\textit{Discussion.}}---%
In this work, we demonstrated that the average two-qubit gate count in a pure quantum circuit can be significantly reduced by approximating the circuit with a mixed quantum channel, wherein gates nearest to Clifford unitaries are stochastically either discarded or replaced by ``overrotated'' gates.
We introduce an automated protocol, which leverages shots naturally occurring on quantum computing architectures, where instead of repeating the same circuit, one runs a collection of slightly different circuits.
This approximation is performed under a strict error budget, which we saturate using an exact analytical expression for the diamond distance of individual gate replacements; 
our results remain the same for experimentally relevant typical distance measures, subject to a multiplicative factor.
Our findings reveal modest reductions in gate count for random circuits, and markedly greater savings for quantum Fourier transform circuits, competing with well-known approximation algorithms.
Moreover, our protocol is compatible with many other approximation algorithms and approximate circuit synthesis protocols, making it a natural, fully automated addition to quantum compilation pipelines.

Our results open several directions for further investigation. 
First, allowing for mixed non-unitary channels could potentially reduce the approximation errors further. However, implementing this in practice may not be beneficial, as measurements tend to be costly on many architectures.
Second, a broader benchmarking effort across diverse circuit architectures is warranted to identify regimes where mixed-channel protocols offer the greatest benefit -- particularly when dealing with frequent small-angle phase gates. Such benchmarks could also uncover structural patterns that allow for new approximation strategies.
Moreover, an interesting direction for refinement would be to introduce dynamic $p$ (either as a function of system-specific variables or the structural circuit neighborhood); or to incorporate device-specific gate errors into the error budget. 
Additionally, our protocol establishes a general paradigm (replacing near-Clifford gates with carefully chosen mixtures) that can be readily applied to other scenarios, such as single-qubit controlled rotations, illustrating the broad utility of the underlying principle.
These enhancements could pave the way for more tailored automatic optimization strategies that combine mixed-channel approximations with structural simplification, and balancing fidelity and efficiency in ways unattainable through pure unitary channels alone.

\begin{acknowledgments}
    M.~S.\@, A.~K.\@, and N.~L.\@ were supported by the Engineering and Physical Sciences Research Council grant on Robust and Reliable Quantum Computing (RoaRQ), Investigation 004 grant reference EP/W032635/1.
    M.~S.\@ and A.~K.\@ were also supported by the Engineering and Physical Sciences Research Council grant reference EP/Z002230/1: (De)constructing quantum software (DeQS).
    The authors would like to acknowledge the use of the University of Oxford Advanced Research Computing (ARC) facility in carrying out this work. 
    \url{http://dx.doi.org/10.5281/zenodo.22558}.
\end{acknowledgments}

\bibliography{refs}


\appendix

\clearpage

\onecolumngrid

\section{End Matter}

\twocolumngrid

\textbf{\textit{Diamond distance of one replacement.}}---%
Here we show the full derivation of the diamond distance between $\mathcal{Z}_\alpha$ and $\mathcal{E}_{\theta, p}$ defined in Eqs.~\eqref{eq:channel_original} and \eqref{eq:channel_approximate}. We have
\begin{align}
  d_{\diamond} (\mathcal{Z}_{\alpha}, \mathcal{E}_{\theta, p}) &= \max_M \| (\mathcal{Z}_{\alpha} \otimes \mathbb{I}) (M) {-} (\mathcal{E}_{\alpha,p} \otimes \mathbb{I}) (M) \|_1,\\
  &=\max_M \left\| \left({\small\begin{array}{cccc}
  0& 0 & B M_{13} & B M_{14}\\
  0& 0 & B M_{23} & B M_{24}\\
  B^{\ast} M_{13}^{\ast} & B^{\ast} M_{23}^{\ast} & 0 & 0\\
  B^{\ast} M_{14}^{\ast} & B^{\ast} M_{24}^{\ast} & 0 & 0
\end{array}}\right) \right\|_1,
\end{align}
where $B = e^{- i \alpha} - e^{- i \theta} (1 - p) - p$, and $M_{ij}$ are matrix elements of $M$. The trace norm $\|A\|_1$ is a sum of singular values of $A$, and hence
\begin{align}
  d_{\diamond} (\mathcal{Z}_{\alpha}, \mathcal{E}_{\theta, p}) &= \sqrt{2} |B| \max_M \left(\sqrt{C-\sqrt{C^2-4 |D|^2}}\right.\\
  &\quad+\left.\sqrt{C+\sqrt{C^2-4 |D|^2}}\right),\nonumber
\end{align}
where $C=|M_{13}|^2+|M_{14}|^2+|M_{23}|^2+|M_{24}|^2\ge 0$ and $D=M_{14}M_{23}-M_{13}M_{24}$. Since $\mathcal{Z}_{\alpha} - \mathcal{E}_{\theta, p}$ is a Hermitian-preserving map, we can write $M=\ket{u}\bra{u}$~\cite{Watrous2018}, where $\ket{u}$ is a normalized vector, which leads to $D=0$, and
\begin{align}
  d_{\diamond} (\mathcal{Z}_{\alpha}, \mathcal{E}_{\theta, p}) &= 2 |B| \max_M \sqrt{C}.
\end{align}
Maximizing $\sqrt{C}$ can be done by maximizing $C=(|u_1|^2+|u_2|^2)(|u_3|^2+|u_4|^2)=x(1-x)$, where $x=|u_1|^2+|u_2|^2$, $u_i$ are matrix elements of $\ket{u}$, and we have used normalization of $\ket{u}$. Function $x(1-x)$ has a maximum at $x=1/2$, so $\max_M \sqrt{C}=1/2$, and
\begin{align}
  d_{\diamond} (\mathcal{Z}_{\alpha}, \mathcal{E}_{\theta, p}) &= |B| = |e^{- i \alpha} - e^{- i \theta} (1 - p) - p|.
\end{align}
Converting this to real functions, we obtain Eq.~\eqref{eq:dist}.

\textbf{\textit{Typical distance measures.}}---%
Here, we calculate several channel distances that capture typical errors, and show that they are proportional to the diamond distance.

First, we leverage the Frobenius norm (also known as the Hilbert-Schmidt norm), $\| A \|_F = \sqrt{\sum_i \sum_j | A_{i j} |^2}$. The average distance can then be measured by averaging the Frobenius distance over random Haar states,
\begin{align}
    d_F(\mathcal{E}_1,\mathcal{E}_2) &= \mathbb{E}[ \| \mathcal{E}_1(\ket{\psi}\bra{\psi}) - \mathcal{E}_2(\ket{\psi}\bra{\psi}) \|_F ],
\end{align}
where $\ket{\psi}$ is a Haar-random state and $\mathbb{E}[\cdot]$ is an ensemble average. For one $Z$-phase replacement, the average Frobenius distance can be written explicitly by parametrizing a random state $| \psi \rangle = \text{CUE} (2) \cdot \{ 1, 0 \} = \left\{ e^{i \varphi_1} \sqrt{a}, e^{i \varphi_2} \sqrt{1 - a} \right\}$, where $a \in [0, 1], \varphi_1 \in [0, 2 \pi), \varphi_2 \in [0, 2 \pi)$ are uniformly distributed. This leads to
\begin{align}
  d_F (\mathcal{Z}_{\alpha}, \mathcal{E}_{\theta, p}) & = \mathbb{E} \left[ \sqrt{a (1 - a)} \right] \Big(| e^{i \alpha} - (1 - p) e^{i \theta} - p |^2 \nonumber \\
  &\quad+ | e^{i \alpha} - (1 - p) e^{i \theta} - p e^{i (\alpha + \theta)} |^2\Big)^{1/2} \\
  & = \frac{\pi}{4 \sqrt{2}} d_{\diamond} (\mathcal{Z}_{\alpha}, \mathcal{E}_{\theta, p}) \\
  & \approx 0.555\ d_{\diamond} (\mathcal{Z}_{\alpha}, \mathcal{E}_{\theta, p}) .
\end{align}
Similarly, one could define the average trace norm distance between channels,
\begin{align}
    d_{1,\text{av}}(\mathcal{Z}_{\alpha},\mathcal{E}_{\theta, p}) &= \mathbb{E} [\| \mathcal{Z}_{\alpha} (| \psi \rangle \langle \psi |) - \mathcal{E}_{\theta, p} (| \psi \rangle \langle \psi |) \|_1] \\
    &= \mathbb{E} \left[ \sqrt{a (1 - a)} \right] 2 | B | \\
    &= \frac{\pi}{4} d_{\diamond} (\mathcal{Z}_{\alpha}, \mathcal{E}_{\theta, p})\\
    &\approx 0.785\ d_{\diamond} (\mathcal{Z}_{\alpha}, \mathcal{E}_{\theta, p}) .
\end{align}
However, both distances are induced distances, and hence, may not obey the composition rules of the tensor product. Instead, it is more advantageous to define the average-case distance between channels~\cite{Maciejewski2023ieee, Maciejewski2023quantum},
\begin{equation}
    d_\text{av}(\mathcal{E}_1, \mathcal{E}_2) = \frac{1}{2} \sqrt{\| \mathcal{J}_{\mathcal{E}_1} {-}\mathcal{J}_{\mathcal{E}_2} \|_F^2 + \text{tr} [(\mathcal{E}_1 (\tau_d) - \mathcal{E}_2 (\tau_d))^2]},
\end{equation}
where $\mathcal{J}_{\mathcal{E}} = \frac{1}{d} \sum_{i, j = 1}^d | i \rangle \langle j | \otimes \mathcal{E} (| i \rangle \langle j |)$ is the Choi-Jamio{\l}kowski state (note the normalization), and $\tau_d =\mathbb{I}_d / d$ is the maximally mixed state, with $d$ being the Hilbert space dimension.
$d_{\text{av}}$ obeys several useful properties, such as the triangle inequality and subadditivity w.r.t.\@ tensor products, allowing us to easily compose multiple replacements in the circuit to calculate the bound on the average distance for the whole circuit.

For one $Z_\alpha$ replacement (where $d = 2$), we find that
\begin{align}
  \mathcal{Z}_{\alpha} (\tau_d) - \mathcal{E}_{\theta, p} (\tau_d) & = 0, \\
  \mathcal{J}_{\mathcal{Z}_{\alpha}} -\mathcal{J}_{\mathcal{E}_{\theta, p}} & = \frac{1}{2}
  \left(\begin{array}{cccc}
    0 & \ 0 \  & \ 0 \  & B\\
    0 & 0 & 0 & 0\\
    0 & 0 & 0 & 0\\
    B^* & 0 & 0 & 0
  \end{array}\right).
\end{align}
This yields
\begin{align}
  d_{\text{av}} (\mathcal{Z}_{\alpha}, \mathcal{E}_{\theta, p}) & = \frac{1}{2 \sqrt{2}} |B| \\
  & = \frac{1}{2 \sqrt{2}} d_{\diamond} (\mathcal{Z}_{\alpha}, \mathcal{E}_{\theta, p}) \\
  & \approx 0.354\ d_{\diamond} (\mathcal{Z}_{\alpha}, \mathcal{E}_{\theta, p}) . 
\end{align}

\begin{figure}[tb]
  \centering
  \includegraphics[width=\columnwidth]{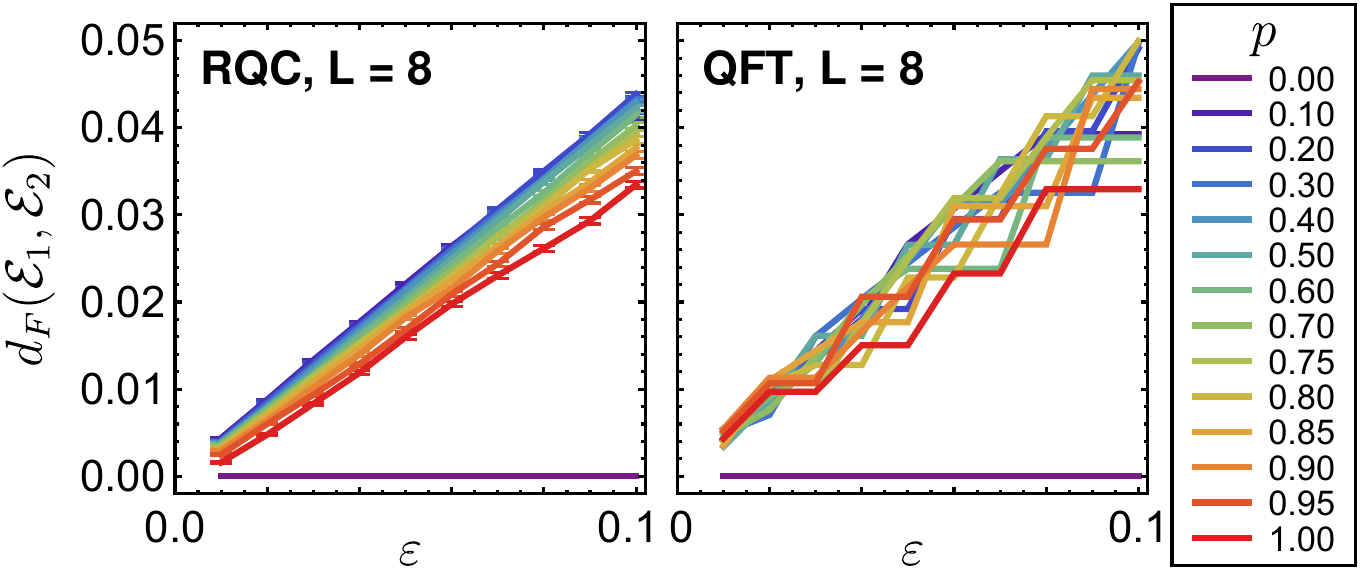}
  \caption{The average Frobenius distance for the random quantum circuit (left) and the quantum Fourier transform (right), both for $L=8$. \label{fig:frobenius_distance}}
\end{figure}

To summarize, for our replacement procedure, we find that several average distance measures are proportional to the diamond distance. Hence, when calculating the diamond distance, which is a useful measure of error from the quantum information theory perspective, we also obtain measures of typical errors, which are more relevant experimentally on quantum hardware.
Numerically, we find that when many replacements are made, the average Frobenius distance between the original and approximate circuit is found to be roughly independent of $p$, and linear in the error budget $\varepsilon$ (see Fig.~\ref{fig:frobenius_distance}). This again agrees with our expectation that typical distance measures are proportional to the diamond distance.

\end{document}